\documentclass[showpacs,preprint,showkeys,nofootinbib,tightenlines]{revtex4}
\usepackage{epsfig}
\usepackage{amsmath}
\usepackage{bbm}
\usepackage{graphicx}
\usepackage{braket}
\usepackage{bm}
\usepackage{color}
\usepackage[T1]{fontenc}
\usepackage{fouridx}
\usepackage{braket}
\usepackage{bm}
\usepackage{color}
\usepackage[T1]{fontenc}
\usepackage{appendix}
\usepackage{verbatim}
\usepackage{multirow}
\usepackage[colorlinks,linkcolor=blue]{hyperref}
\usepackage{subfigure}
\usepackage{slashed}
\usepackage{textcomp}
\begin{document}
%\begin{CJK*}{UTF8}{gbsn}

%\title{Next-to-leading order corrections to inclusive $h_c$ production at $B$ factories}
%\title{Next-to-leading order QCD corrections to $ H\to J/\psi(\Upsilon) + Z $\footnote{Supported by National Natural Science Foundation of China (11375168).}}

%\title{Next-to-leading order QCD corrections to the decay of Higgs to vector meson and $ Z $ boson\footnote{Supported by National Natural Science Foundation of China (11375168).}}

\title{Next-to-leading order QCD corrections to the decay of Higgs to vector meson and $ Z $ boson}

\author{Qing-Feng Sun\footnote{qfsun@mail.ustc.edu.cn}}
\affiliation{Department of Modern Physics, University of Science and Technology of China, Hefei, Anhui 230026, China}
\author{An-Min Wang}
\affiliation{Department of Modern Physics, University of Science and Technology of China, Hefei, Anhui 230026, China}
%\date{\today}
\begin{abstract}
The exclusive decay of the Higgs boson to a vector meson ($ J/\psi $ or $ \Upsilon(1S) $) and $ Z $ boson is studied in this work. The decay amplitudes are separated into two parts in a gauge invariant manner. The first part comes from the direct coupling of the Higgs boson to the charm (bottom) quark and the other  from the $ HZZ^{*} $ or the loop-induced $ HZ\gamma^{*} $ vertexes in the standard model. While the branching ratios from the direct channel are much smaller than those of the indirect channel, their interference terms give nontrivial contributions. We further calculate the QCD radiative corrections to both channels, which reduce the total branching ratios by around 20\% for both $ J/\psi $ and $ \Upsilon(1S) $ production. These results may help to check the SM predictions of the $ H c\bar{c}(H b\bar{b}) $ coupling and to seek for hints of new physics at the High Luminosity LHC or future hadron colliders.


\end{abstract}

\maketitle

\section{Introduction}
The discovery of the 125 GeV Higgs-like boson by the ATLAS and CMS collaborations~\cite{Aad:2012tfa,Chatrchyan:2012xdj} has attracted much interest from both experimental and theoretical aspects. It is most important to precisely determine the properties of this new boson to check the predictions of standard model (SM). No significant deviation from the SM has been found from measurements of the decays $ H\to \gamma\gamma ,ZZ\text{ and } WW$~\cite{ATLAS:2013sla,CMS:yva}.
%Its coupling to gauge bosons and the third generation fermions are measured directly~\cite{ATLAS:2013sla,CMS:yva} and no significant deviation from the predictions of SM has been found.

In the SM, the coupling of the Higgs to the first and second generation fermions is very weak and thus difficult to measure directly. In Ref.~\cite{Bodwin:2013gca}, the authors point out that the $ Hc\bar{c} $ coupling may be probed by the measurements of the $ H\to J/\psi + \gamma $ decay mode. Their results show that the partial width via the direct decay channel, where the Higgs boson is coupled directly to the charm quark, is too small to be probed. They introduce a new mechanism where the Higgs boson decays through the loop-induced $ H\gamma \gamma^{*} $ vertex, followed by $ \gamma^{*}\to J/\psi $. This indirect mechanism turns out to dominate the $ H\to J/\psi +\gamma $ decay, and the interference between the direct and the indirect channel may provide us with more detailed information about the $ Hc\bar{c} $ coupling.

The proposed High Luminosity LHC (HL-LHC) project is designed to run at 14~TeV, with its integrated luminosity upgraded to $ 3000~\text{fb}^{-1} $. The HL-LHC will allow us to perform precise measurements that are impossible otherwise, which motivates us to make a more detailed study of the rare decays of the SM Higgs boson. The small Yukawa coupling of the light fermions leaves space for theories of new physics beyond the SM and has attracted a lot of interest. In Refs.~\cite{Gonzalez-Alonso:2014rla,Gao:2014xlv,Modak:2014ywa,Bhattacharya:2014rra,Alte:2016yuw}, the rare decay $ H\to V +Z $ ($ V=J/\psi \text{ or } \Upsilon(1S) $) has been studied. Very similar to $ H\to V + \gamma $, we can separate the decay amplitudes into two parts. The first part comes from the direct coupling of Higgs boson to the charm (bottom) quark, and the other part from the $ HZZ^{*} $ or the loop-induced $ HZ\gamma^{*} $ vertexes in the SM, followed by $ Z^{*}(\gamma^{*})\to V $. Since both the vector meson and $ Z $ boson can decay to a pair of charged leptons, this experimentally-clean final state may provide us a chance to probe this rare decay and seek for a hint of new physics.

%The leading order(LO) results for $ H\to V + Z $ in Ref.~\cite{Gao:2014xlv}  show that the decay width via the loop-induced $ H\to Z\gamma^{*} $ vertex is comparable with the contribution from the tree level $ HZZ^{*} $ vertex in SM. 
In Ref.~\cite{Gao:2014xlv}, the author shows that the decay rate of $ H\to V+Z $ via the loop-induced $ HZ\gamma^{*} $ vertex is comparable with the contributions from the tree level $ HZZ^{*} $ vertex in the SM, by analyzing this decay via the indirect channel at leading-order (LO). In Ref.~\cite{Modak:2014ywa}, the authors analyze the decay $ H\to V+Z $ from all channels, and interference contributions are also included. They further show that this rare decay is sensitive to the anomalous Higgs couplings originating from physics beyond the SM, by observations of the decay of the final-state vector meson and $ Z $ boson into charged leptons. In this paper, we revisit this rare decay at LO, and evaluate its QCD corrctions at next-to-leading order (NLO) for the (un)polarized final-state vector meson and $ Z $ boson. We also carefully analyze the interference terms between the indirect channel and the direct one at both LO and NLO level.
%*******************************************************************%

This paper is organized as follows. In Section~\ref{Sec:formalism and notations}, we introduce the formalism and notations used in this work. In Section~\ref{Sec:QCD corrections}, we revisit the decay $ H\to V+Z $ at LO, and further we calculate its QCD corrections at NLO. In Section~\ref{Sec:polarized results}, we analyze the helicity amplitudes by projecting the vector meson and $ Z $ boson to particular polarized states. We conclude our calculations in Section~\ref{Sec:conclusion}.

\section{Formalism and notations}\label{Sec:formalism and notations}
	
	For the $ H\to V+Z $ decay, there are several distinct dynamic scales: $ m_H\sim m_Z> m_q $  $ (m_q=m_b \text{ or } m_c) $. There is another scale in heavy quarkonium production, $ m_q v_r $, where $ v_r $ is the relative velocity of the heavy quark pair in the rest frame of the quakonium. For heavy quarkonium, $ v_r\ll 1 $ is assumed, making it a non-relativistic system. The scale $ m_q v_r $ characterizes the hadronization process of the heavy quark pair to vector mesons, which is inherently nonperturbative. 
	
	We adopt the nonrelativistic QCD (NRQCD) factorization formula as described in Ref.~\cite{Bodwin:1994jh}. In the NRQCD factorization framework, the production of heavy mesons can be separated into two parts. First, the constituent quarks are produced at an energy greater than the heavy quark mass and thus can be calculated perturbatively. Secondly, the quark pair binds into quarkonium at an energy much smaller than the heavy quark mass, and this process is represented by the non-perturbative long distance matrix elements (LDMEs). In this formula, the LDMEs can be expanded by the relative velocity $ v_r $ of the heavy quark pair and the short-distance coefficients are expanded perturbatively by the strong coupling $ \alpha_s $ order by order. As a result, the production rate of the heavy quarkonium can be expressed as the sum of products of the non-perturbative LDMEs and the corresponding perturbative short-distance coefficients. We will expand the short-distance coefficients to NLO of the strong coupling $ \alpha_s $ and the LDMEs are expanded to the lowest order of $ v_r $ throughout this work.
%	\par
%	In the NRQCD frame work, the amplitude for $ H\to J/\psi+Z $ can be expressed as\footnote{The formula for $ \Upsilon $ is similar.}
%	\begin{equation}\label{NRQCD formula}
%	\mathcal{M}\left[H\to J/\psi(\lambda)+Z\right]=\sqrt{2M_{J/\psi}}\left[c_0(\lambda)\braket{J/\psi|\psi^{\dagger}\bm{\sigma}\cdot\bm{\epsilon}\chi|0}+\frac{c_{2}(\lambda)}{m_c^2}\braket{J/\psi|\psi^{\dagger}\bm{\sigma}\cdot\bm{\epsilon}\left(-\frac{i}{2}\overleftrightarrow{\bm{D}}\right)^2\chi|0}+...\right],
%	\end{equation}
%	where in Eq.~\ref{NRQCD formula}, the first and second terms in the square brackets of right hand represent the order-$ v_r^0 $ and order-$ v_r^2 $ corrections respectively, and $ c_i(\lambda)(i=1,2) $ represent the corresponding short-distance coefficients. In this work, we only retain the order-$ v_r^0 $ term and expand the coefficient $ c_0(\lambda) $ to NLO by the strong coupling $ \alpha_s $ perturbatively.
%($ J/\psi \text{ or } \Upsilon(1S) $)

We define the momenta of the initial-state Higgs and the outgoing vector meson and $ Z $ boson as
\begin{equation}
H(p_H)\to V(p,\lambda_V)+Z(k,\lambda_Z),
\end{equation}
where $ p= p_q+p_{\bar{q}} $ and $ k $ represent the momenta of the final-state vector meson and $ Z $ boson respectively, with
\begin{equation}
p = \left(\sqrt{m_V^2+|\bm{p}|^2},\bm{p}\right), \qquad k = \left(\sqrt{m_Z^2+|\bm{p}|^2},-\bm{p}\right),
\end{equation}
and
\begin{equation}
|\bm{p}|= \frac{1}{2m_H}\sqrt{m_H^4+m_Z^4+m_V^4-2m_H^2 m_V^2-2m_H^2 m_Z^2-2m_V^2 m_Z^2}.
\end{equation}
$ m_V $ is the mass of the vector meson and $ m_V= 2m_q $ at the lowest order of $ v_r $. $ p_H= p+k $ is the momentum of the initial-state Higgs boson and $ \lambda_V(\lambda_Z)=0,\pm 1 $ represent the helicities of the final-state vector meson ($ Z $ boson).

To project the free quark pair into our desired $ \fourIdx{3}{}{}{1}{S} $ and color singlet quantum state, we adopt the relatively normalized spin projection operator in Eq.~(A9b) of Ref.~\cite{Bodwin:2010fi}:
\begin{equation}\label{spin projection}
\Pi_V=\frac{1}{4\sqrt{2}E_q\left(E_q+m_q\right)}(\slashed{p}_{\bar{q}}-m_q)\slashed{\epsilon}_V^{*}(p,\lambda_V)(\slashed{p}+2E_q)(\slashed{p}_q+m_q)\otimes \frac{\bm{\mathbbm{1}}}{\sqrt{N_c}},
\end{equation}
where $ E_q $ in Eq.~\eqref{spin projection} is the energy of the constituent quarks in the rest frame of the heavy quark pair. Since we expand the LDMEs to the lowest order of $ v_r $, the relative momentum between the quark pair is thus neglected. As a result, the momenta of the constituent heavy quarks are assigned as
\begin{equation}
p_q=p_{\bar{q}}=\frac{1}{2} p,
\end{equation}
and $ E_q= m_q $.

By Lorentz covariance, the amplitudes for $ H\to V + Z $ can be decomposed as
\begin{align}\label{helicity amplitudes}
\mathcal{M}_{\lambda_V\lambda_Z}\left[H\to V+Z\right]=&\mathcal{T}_{\mu\nu}\epsilon_{V}^{\mu *}(p,\lambda_V)\epsilon_{Z}^{\nu *}(k,\lambda_Z)
\notag\\
=&F_1 \epsilon_{V}^{*}(p,\lambda_V)\cdot\epsilon_{Z}^{*}(k,\lambda_Z)+F_2 k\cdot \epsilon_{V}^{*}(p,\lambda_V) p\cdot \epsilon_{Z}^{*}(k,\lambda_Z),
\end{align}
where $ \epsilon_{V}^{\mu}(p,\lambda_V) $ and $ \epsilon_{Z}^{\nu}(k,\lambda_Z) $ are the polarization vectors of the final-state vector meson and $ Z $ boson, and their explicit expressions will be given in Section~\ref{Sec:polarized results}. We will show in Section~\ref{Sec:polarized results} that the helicity amplitudes in Eq.~\eqref{helicity amplitudes} are free of polar and azimuthal angles.

The decay width of $ H\to V + Z $ can be expressed as
\begin{equation}\label{Eq:decay width}
\Gamma(H\to V + Z)=\frac{|\bm{p}|}{8\pi m_H^2}\sum_{pol}\left|\mathcal{M}_{\lambda_V\lambda_Z}\left[H\to V + Z\right]\right|^2.
\end{equation}
At order-$ v_r^0 $, the helicity amplitudes in Eq.~(\ref{Eq:decay width}) can be expressed as
\begin{align}\label{Eq:NRQCD amplitude}
\mathcal{M}_{\lambda_V\lambda_Z}\left[H\to V + Z\right]=&\sqrt{\frac{2m_V\braket{\mathcal{O}_1}_V}{2N_c(2E_q)^2}}\mathcal{M}_{\lambda_V\lambda_Z}\left[H\to q\bar{q}+Z\right]
\notag\\
=&\sqrt{\frac{\braket{\mathcal{O}_1}_V}{2N_c m_c}}\mathcal{M}_{\lambda_V\lambda_Z}\left[H\to q\bar{q}+Z\right].
\end{align}
%\begin{equation}\label{Eq:NRQCD amplitude}
%\mathcal{M}_{\lambda_V\lambda_Z}\left[H\to V + Z\right]=\sqrt{\frac{2m_V\braket{\mathcal{O}_1}_V}{2N_c(2E_q)^2}}\mathcal{M}_{\lambda_V\lambda_Z}\left[H\to q\bar{q}+Z\right],
%\end{equation}
where $ \braket{\mathcal{O}_1}_V $ represents the nonperturbative LDME and we take $ \braket{\mathcal{O}_1}_{J/\psi}= 0.44~\text{GeV}^3 $ and $\braket{\mathcal{O}_1}_{\Upsilon(1S)}=3.07~\text{GeV}^3 $ for the case of $ J/\psi $ and $ \Upsilon (1S)$, respectively~\cite{Bodwin:2007fz}.

For the phenomenological results, we take $ m_H= 125.09~\text{GeV}$, $ m_Z=91.1876~\text{GeV} $, $ m_W=80.385~\text{GeV} $, and $ m_t=m_t \text{(pole mass)}=174.2~\text{GeV} $~\cite{Patrignani:2016xqp}. The decay width of the Higgs boson $ \Gamma_H=4.100~\text{MeV} $~\cite{deFlorian:2016spz}. The electroweak coupling is parameterized as $ \alpha_{G_\mu}=\frac{\sqrt{2}}{\pi}G_{\mu}m_W^2\left(1-m_W^2/m_Z^2\right) \approx 1/132.23 $, and the Fermi coupling constant $ G_{\mu}=1.1663787\times 10^{-5}~\text{GeV}^{-2} $. The strong coupling $ \alpha_s(m_H/2) \approx 0.1253 $, which is evaluated with $ \textsf{RunDec} $~\cite{Herren:2017osy}. We take $ m_c=1.5~\text{GeV} $ and $ m_b=4.6~\text{GeV} $ for the masses of the charm and bottom quarks, respectively.
\section{QCD corrections to unpolarized $ H\to V+Z $}\label{Sec:QCD corrections}
\subsection{LO results revisited}
\begin{figure}[htbp]
	\centering
	\includegraphics[width=0.8\linewidth]{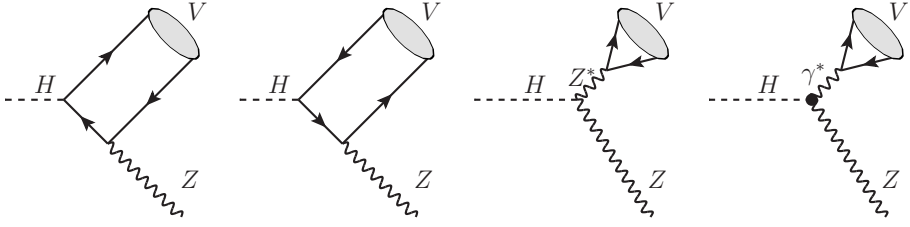}
	\caption{Feynman diagrams for LO $ H\to V + Z $. The Feynman diagrams were generated with \textsf{JaxoDraw}~\cite{Binosi:2003yf}.}
	\label{fig:lo}
\end{figure}

Feynman diagrams for LO $ H\to V + Z $ are shown in Fig.~\ref{fig:lo}. The first two diagrams come from the direct channel where the charm (bottom) quark is coupled directly to the Higgs boson. The last two diagrams come from the $ HZZ^* $ and the loop-induced $ HZ\gamma^* $ vertexes in the SM, respectively.

\begin{table}[htbp]
	\renewcommand\arraystretch{1.2}
	\begin{tabular}{|c|c|c|c|c|}
		\hline 
		& direct & indirect & interference & total \\ 
		\hline 
		$ J/\psi $ & $ 8.34\times 10^{-10} $ & $ 4.36\times 10^{-6} $ & $ -6.96\times 10^{-8} $ & $ 4.30\times 10^{-6} $ \\
		\hline 
		$ \Upsilon(1S) $ & $ 5.22\times 10^{-8} $ & $ 2.14\times 10^{-5} $ & $ 8.43\times 10^{-7} $ & $ 2.23\times 10^{-5} $ \\  
		\hline 
	\end{tabular}
	\caption{Branching ratios for $ H\to V + Z $ at LO.}
	\label{Tab:LO results}
\end{table}
In Table~\ref{Tab:LO results}, we list the branching ratios for $ H\to V + Z $ from the direct, the indirect and the interference channels\footnote{In Ref.~\cite{Modak:2014ywa}, the authors list the LO results in their Table~1, and we find some disagreements with their results. The expression $ m_H^2-m_Z^2-m_V^2 $ in the denominator of Eq.~(6) in Ref.~\cite{Modak:2014ywa} should be $ m_H^2+m_Z^2-m_V^2 $.}. The branching ratios are rather small at LO for both $ J/\psi $ and $ \Upsilon(1S) $. The contributions from the direct channels of $ J/\psi $ and $\Upsilon(1S) $ production are totally negligible, as expected. The branching ratios from the indirect channel dominate the total results and the interference branching ratios amount to -1.60\% and 3.93\% of the total results for $ J/\psi $ and $ \Upsilon(1S) $ respectively.
\subsection{QCD corrections}
\begin{figure}[htbp]
	\centering
	\includegraphics[width=0.8\linewidth]{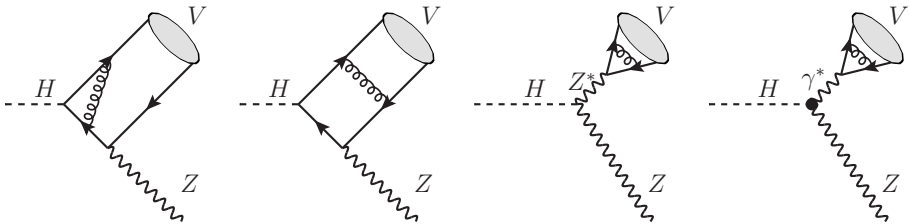}
	\caption{Representative Feynman diagrams for QCD corrections to $ H\to V + Z $.}
	\label{fig:oneloop}
\end{figure}

QCD radiative corrections are performed by attaching gluons to the final-state quarks as shown in Fig.~\ref{fig:oneloop}. Note that the loop-induced $ HZ\gamma^* $ vertex may also have QCD corrections at NLO. In Refs.~\cite{Bonciani:2015eua,Gehrmann:2015dua}, the authors calculate the QCD corrections to the rare decay $ H\to Z \gamma $ at NLO and the relative corrections turn out to be rather small, around 3\textperthousand\ of the LO decay width. We neglect the tiny QCD corrections to the $ H Z \gamma^* $ vertex throughout this work and extract the parameters for the effective $ HZ\gamma^* $ coupling by evaluating this vertex at LO.

We choose the dimensional regulation to regularize the possible UV and IR singularities in $ d=4-2\epsilon $ dimensions. The renormalization constants $ Z_2 $ and $ Z_m $ are defined as
\begin{equation}
\psi_q^0=\sqrt{Z_2}\psi_q, \qquad m_q^0=Z_m m_q,
\end{equation}
where $ \psi_q(\psi_q^0) $ and $ m_q(m_q^0) $ represent the renormalized(bare) quark field and mass, respectively. In the on-shell (OS) scheme, the renormalization constants expanded to $ O(\alpha_s) $ can be expressed as
\begin{subequations}
	\begin{align}
	Z_2^{\text{OS}}&=1-C_F \frac{\alpha_s}{4\pi}\left(\frac{1}{\epsilon_{\text{UV}}}+\frac{2}{\epsilon_{\text{IR}}}-3\gamma_E+3\ln{\frac{4\pi \mu^2}{m_q^2}}+4\right)+\mathcal{O}\left(\alpha_s^2\right),
	\\
	Z_m^{\text{OS}}&=1-3C_F \frac{\alpha_s}{4\pi}\left(\frac{1}{\epsilon_{\text{UV}}}-\gamma_E+\ln{\frac{4\pi \mu^2}{m_q^2}}+\frac{4}{3}\right)+\mathcal{O}\left(\alpha_s^2\right).
	\end{align}
\end{subequations}

The singularities are removed after renormalization, and the decay width for $ H\to V + Z $ at NLO can be expressed as
%$$
%\Gamma_H^{\text{NLO}}=\Gamma_H^{\text{LO}}(1+2\delta_{\text{QCD}}).
%$$
$$
\Gamma^{\text{NLO}}\equiv \Gamma^{\text{LO}}(1+2\delta_V).
$$
For the QCD corrections to the direct decay channel, $ \delta_{J/\psi}^{\text{dir}}= -60.98\%$ and $ \delta_{\Upsilon(1S)}^{\text{dir}}= -48.88\%$ for $ J/\psi $ and $ \Upsilon(1S) $ respectively. We can see that the results for the direct channel are greatly reduced by the NLO QCD corrections, which makes the NLO results for the direct channel of $ H\to J/\psi +Z $ negative. In Ref.~\cite{Chao:2016usd}, the authors evaluated the NLO corrections to $ H\to V+\gamma $ and the large and negative corrections to the direct channel are similar. At scale $ \mu=m_H/2 $, the relative corrections $ \Delta \Gamma^{\text{direct}}/\Gamma^{\text{direct}} $ defined in Ref.~\cite{Chao:2016usd} are around -85.38\% and -68.38\% for  $ J/\psi $ and $ \Upsilon(1S) $ respectively. These large and negative corrections at NLO may be compensated by the NNLO or the relativistic contributions. The QCD radiative corrections to the indirect channel come from the last two diagrams in Fig.~\ref{fig:oneloop}. The results are trivial and independent of the quark species:
\begin{equation}
\delta_V^{\text{ind}}=-2C_F\frac{\alpha_s}{\pi}\approx -10.63\%.
\end{equation}

\begin{table}[htbp]
	\renewcommand\arraystretch{1.2}
	\begin{tabular}{|c|c|c|c|c|c|}
		\hline 
		& direct & indirect & interference & total \\ 
		\hline 
		$ J/\psi $ & $ -1.83\times 10^{-10} $ & $ 3.44\times 10^{-6} $ & $ -1.87\times 10^{-8} $ & $ 3.42\times 10^{-6} $ \\
		\hline 
		$ \Upsilon(1S) $ & $ 1.17\times 10^{-9} $ & $ 1.69\times 10^{-5} $ & $ 3.61\times 10^{-7} $ & $ 1.72\times 10^{-5} $ \\ 
		\hline 
	\end{tabular}
	\caption{Branching ratios for $ H\to V + Z $ at NLO.}
	\label{Tab:NLO results}
\end{table}
\begin{figure}[htbp]
	\centering
	\includegraphics[width=0.7\linewidth]{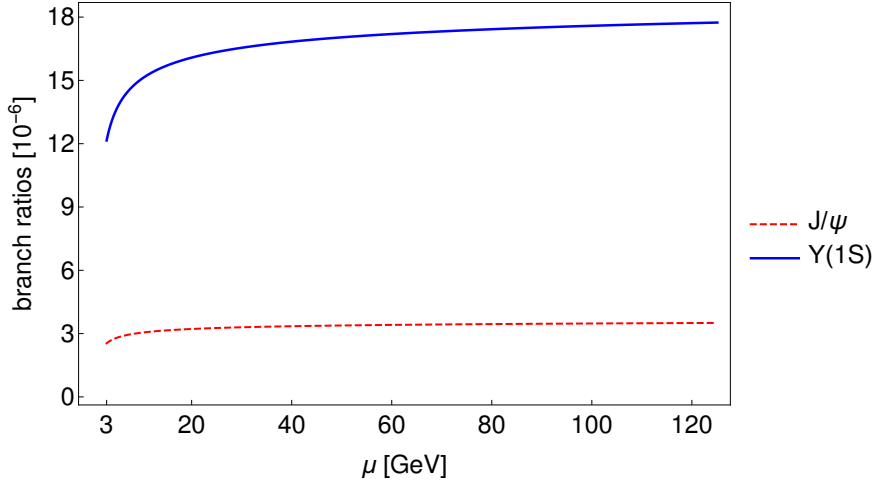}
	\caption{The dependence of the branching ratios for the decay $ H\to V+Z $ on the scale $ \mu $ at NLO. The scale $ \mu $ stems from the QCD running coupling $ \alpha_s(\mu) $, which is taken from $ 2m_c=3.0~\text{GeV} $ to $ m_H=125.09~\text{GeV} $. We evaluate the running coupling $ \alpha_s(\mu) $ with the package \textsf{RunDec} and the initial value of $ \alpha_s(\mu) $ is taken as $ \alpha_s(m_Z)=0.1181 $.}
	\label{fig:scaledependence}
\end{figure}
The NLO results for $ H\to V+Z $ are listed in Table~\ref{Tab:NLO results}. Both of the branching ratios from the direct and the interference channels are greatly reduced, which makes these results sensitive to the strong coupling $ \alpha_s $. In Fig.~\ref{fig:scaledependence}, we show the dependence of the total branch ratios on the scale by sliding $ \mu $ from $ 2m_c $ to $ m_H $.

In this work, we generate the decay amplitudes with \textsf{FeynArts}~\cite{Hahn:2000kx}, and employ the the package \textsf{FeynCalc}~\cite{Shtabovenko:2016sxi} to deal with the traces over the Dirac matrices. The amplitudes are further decomposed with the package \textsf{Apart}~\cite{Feng:2012iq} and the IBP (integration by parts) reductions are performed with \textsf{FIRE}~\cite{Smirnov:2014hma}. The master integrals are calculated with \textsf{Package-X}~\cite{Patel:2016fam} analytically and further checked numerically with \textsf{LoopTools}~\cite{Hahn:1998yk}.

\section{Results for polarized vector meson and $ Z $ boson}\label{Sec:polarized results}
It is sometimes useful to project the final-state vector meson and $ Z $ boson to particular helicity states. The corresponding polarization vectors are
\begin{subequations}\label{polarization vectors}
	\begin{align}
	\epsilon_{V}^{\mu}(p,\pm 1)=&\sqrt{\frac{1}{2}}e^{\pm i\phi}(0,\mp \cos{\theta}\cos{\phi}+i\sin{\phi},-i\cos{\phi}\mp\cos{\theta}\sin{\phi},\pm\sin{\theta}),
	\\
	\epsilon_{Z}^{\mu}(k,\pm 1)=&\sqrt{\frac{1}{2}}e^{\mp i\phi}(0,\pm \cos{\theta}\cos{\phi}+i\sin{\phi},-i\cos{\phi}\pm\cos{\theta}\sin{\phi},\mp\sin{\theta}),
	\end{align}
	and
	\begin{align}
	\epsilon_{V}^{\mu}(p,0)=&\frac{\sqrt{m_V^2+|\bm{p}|^2}}{|\bm{p}| m_V}p^{\mu}-\frac{m_V}{|\bm{p}|m_H}(p^{\mu}+k^{\mu}),
	\\
	\epsilon_{Z}^{\mu}(k,0)=&\frac{\sqrt{m_Z^2+|\bm{p}|^2}}{|\bm{p}| m_Z}k^{\mu}-\frac{m_Z}{|\bm{p}|m_H}(p^{\mu}+k^{\mu}),
	\end{align}
\end{subequations}
where we have introduced $ \theta $, $ \phi $ as the polar and azimuthal angles of $ \bm{p} $ with respect to a fixed $ z $-axis. The \textit{Jacob-Wick convention}~\cite{Jacob:1959at} is adopted for the polarization vector of the final-state $ Z $ boson in Eq.~\eqref{polarization vectors}.

Since the initial-state Higgs boson is spin-0, the helicity amplitudes must be proportional to
\begin{equation}
D^{J}_{\lambda_{i}\lambda_{f}}(\theta,\phi)=D^{0}_{00}(\theta,\phi)=P_0(\cos\theta)=1,
\end{equation}
where $  D^{J}_{\lambda_{i}\lambda_{f}}(\theta,\phi) $ is the \textit{Wigner-D} matrix element in the partial wave expansion~\cite{Haber:1994pe}, and $ \lambda_i=0 $, $ \lambda_f=\lambda_V-\lambda_Z=0 $. Thus the helicity amplitudes defined in Eq.~\eqref{helicity amplitudes} are free of the polar and azimuthal angles $ \theta $, $ \phi $, and $ \lambda_V=\lambda_Z=0, \pm1$.
\begin{table}[htbp]
	\renewcommand\arraystretch{1.2}
	\begin{tabular}{|c|c|c|c|c|c|c|}
		\hline 
		\multirow{2}{*}{}& \multicolumn{3}{|c|}{LO} & \multicolumn{3}{|c|}{NLO} \\ 
		\cline{2-7}
		& T & L & total & T & L & total \\ 
		\hline 
		$ J/\psi $ & $ 1.60\times 10^{-6} $ & $ 2.69\times 10^{-6} $ & $ 4.30\times 10^{-6} $ & $ 1.30\times 10^{-6} $ & $ 2.12\times 10^{-6} $ & $ 3.42\times 10^{-6} $ \\ 
		\hline 
		$ \Upsilon(1S) $ & $ 1.85\times 10^{-6} $ & $ 2.05\times 10^{-5} $ & $ 2.23\times 10^{-5} $ & $ 1.22\times 10^{-6} $ & $ 1.60\times 10^{-5} $ & $ 1.72\times 10^{-5} $ \\ 
		\hline 
	\end{tabular}
	\caption{The branching ratios for the polarized vector meson and $ Z $ boson for $ H\to V + Z $. The notations ``T" and ``L" represent the transversely and longitudinally polarized vector meson and $ Z $ boson.}
	\label{Tab:polarized results}
\end{table}

The numerical results for the decay of the Higgs to the polarized vector meson and $ Z $ boson are shown in Table~\ref{Tab:polarized results}. For the case of $ H\to J/\psi+Z $, the branching ratios for the longitudinally polarized $ J/\psi $ are around 62\% of the total branching ratios at LO and NLO. While the branching ratios for the transversely polarized $ \Upsilon(1S) $ and $ J/\psi $ are very close, the results for the longitudinally polarized $ \Upsilon(1S) $ are one order of magnitude larger than those of $ J/\psi $. As a result, the branching ratios for the longitudinally polarized $ \Upsilon(1S) $ dominate the total branching ratios of the decay $ H\to \Upsilon(1S)+Z $ at both LO and NLO.

\section{Conclusion}\label{Sec:conclusion}
In this work, we have revisited the rare decay $ H\to V+Z $ ($ V=J/\psi \text{ or } \Upsilon(1S)$) in the SM. We separated the decay amplitudes into two parts: the first part from the direct Yukawa coupling of the Higgs boson to the charm (bottom) quark, and the second from the $ HZZ^* $ and the loop-induced $ HZ\gamma^* $ vertexes in the SM. We also analyzed their interference effects carefully. We took a further step by evaluating the QCD radiative corrections to both cases at NLO. The QCD corrections reduce the LO branching ratio by around $ 20\% $. Our results show that the contributions from the direct channel are negligible compared with the indirect channel, due to the tiny Yukawa coupling between the Higgs and the charm (bottom) quark. The interference branching ratios turn out to reach several percent level of the total results at both LO and NLO.

The decay of the Higgs to the polarized vector meson and $ Z $ boson was studied in Section~\ref{Sec:polarized results}. For the decay $ H\to J/\psi+Z $, the production of the longitudinally polarized $ J/\psi $ is about twice that of the transversely polarized $ J/\psi $. For the decay $ H\to \Upsilon(1S)+Z $, our results show that the decay to the longitudinally polarized $ \Upsilon(1S) $ dominates.

Although the branching ratios from the indirect and interference terms give nontrivial contributions at both LO and NLO, the measurements of the $ Hc\bar{c} $ and $ Hb\bar{b} $ couplings seem to be rather difficult via the rare decay $ H\to V+Z $. The small Yukawa coupling may be enhanced by new physics beyond the SM, which provide a chance to observe this rare decay and to test the SM predictions in the projected HL-LHC or in future hadron colliders.
\section{Acknowledgments}
This work is supported by the National Natural Science Foundation of China under Grant No. 11375168.

%\bibitem{CMS:yva}
%[CMS Collaboration],
%CMS-PAS-HIG-13-005.

\clearpage

%\end{CJK*}
\end{document}